\documentclass[conference]{IEEEtran}
\IEEEoverridecommandlockouts
\usepackage{cite}
\usepackage{amsmath,amssymb,amsfonts}
\usepackage{algorithmic}
\usepackage{graphicx}
\usepackage{textcomp}

\usepackage{CJKutf8}
\usepackage{xcolor}

\usepackage{multirow}
\usepackage{multicol}

\usepackage{graphicx}
\usepackage{subfigure} 
\usepackage{float}

\usepackage{booktabs}
\usepackage{amsmath}
\usepackage{setspace}
\usepackage{array,caption}
\usepackage[labelfont=bf]{caption}
\usepackage{threeparttable}
\usepackage{tabularx}
\usepackage{hhline}
\usepackage{booktabs}

\usepackage{longtable}

\usepackage{makecell}

\usepackage{hyperref}

\usepackage{pifont}

\usepackage{ragged2e}
\def\BibTeX{{\rm B\kern-.05em{\sc i\kern-.025em b}\kern-.08em
    T\kern-.1667em\lower.7ex\hbox{E}\kern-.125emX}}

\usepackage{float}
\usepackage{booktabs}
\usepackage{multirow}

\begin{document}
\begin{CJK}{UTF8}{gbsn}

\title{An empirical study of next-basket recommendations 
}

\author{
\IEEEauthorblockN{
    Zhufeng Shao\IEEEauthorrefmark{1},
    Shoujin Wang\IEEEauthorrefmark{2},
    Qian Zhang\IEEEauthorrefmark{1},
    Wenpeng Lu\thanks{ \ding{41} Wenpeng Lu is the corresponding author.}\IEEEauthorrefmark{1}\textsuperscript{(\ding{41})},
    Zhao Li\IEEEauthorrefmark{3},
    Xueping Peng\IEEEauthorrefmark{4}
}
\IEEEauthorblockA{
    \IEEEauthorrefmark{1} School of Computer, Qilu University of Technology (Shandong Academy of Sciences), Jinan, China\\ 
    \IEEEauthorrefmark{2} The Data Science Institute, University of Technology Sydney, Sydney, Australia \\
    \IEEEauthorrefmark{3} Shandong Evay Info Technology Co., Ltd., Jinan, China\\Shandong Computer Science Center (National Supercomputer Center in Jinan), Jinan, China  \\
    \IEEEauthorrefmark{4} Australian Artificial Intelligence Institute, University of Technology Sydney, Sydney, Australia \\
    Email: zhufengshao7@gmail.com, shoujin.wang@uts.edu.au, qianzhang9706@gmail.com,\\ lwp@qlu.edu.cn, liz@sdas.org, xueping.peng@uts.edu.au}
}







\maketitle

\begin{abstract}

Next Basket Recommender Systems (NBRs) function to recommend the subsequent shopping baskets for users through the modeling of their preferences derived from purchase history, typically manifested as a sequence of historical baskets. Given their widespread applicability in the E-commerce industry, investigations into NBRs have garnered increased attention in recent years. Despite the proliferation of diverse NBR methodologies, a substantial challenge lies in the absence of a systematic and unified evaluation framework across these methodologies. Various studies frequently appraise NBR approaches using disparate datasets and diverse experimental settings, impeding a fair and effective comparative assessment of methodological performance. To bridge this gap, this study undertakes a systematic empirical inquiry into NBRs, reviewing seminal works within the domain and scrutinizing their respective merits and drawbacks. Subsequently, we implement designated NBR algorithms on uniform datasets, employing consistent experimental configurations, and assess their performances via identical metrics. This methodological rigor establishes a cohesive framework for the impartial evaluation of diverse NBR approaches. It is anticipated that this study will furnish a robust foundation and serve as a pivotal reference for forthcoming research endeavors in this dynamic field.


\end{abstract}
\begin{IEEEkeywords}
recommender systems,next basket recommendation, evaluation
\end{IEEEkeywords}


\section{Introduction}

Over the years, Recommender Systems (RSs) have exhibited noteworthy advancements across diverse real-world applications, including but not limited to e-commerce\cite{qin2021world,qian2022} and streaming media\cite{lu2022aspect,lu2019graph,lu2021sentence}. Their pivotal roles have assumed heightened significance within the contemporary Pan information age, wherein users are confronted with the imperative task of navigating through an exponentially expanding array of information resources to make informed choices. In response to this escalating need, Recommender Systems have emerged as indispensable instruments, enabling users to navigate efficiently and effectively through extensive and often redundant content, products, and services \cite{wang2021survey}. Notably, within the realm of Recommender Systems, a burgeoning area of interest in recent years pertains to next-basket recommender systems (NBRs), signifying a nuanced sub-domain that merits particular attention and investigation.

The examination of Next-Basket Recommender Systems (NBRs) is warranted due to their relevance in practical shopping scenarios. Consumers typically purchase a collection of items, known as a "shopping basket," during a single shopping visit, reflecting authentic shopping patterns. NBRs align with this behavior, aiming to provide recommendations for a carefully selected basket of items based on the user's current needs and preferences. In contrast to prevalent next-item recommendation approaches that focus on the subsequent item within an ongoing basket \cite{song2021next,wang2020modelling,wang2019modeling,wang2018attention}, NBRs address the more comprehensive next-basket recommendation task. This involves recommending a sequence of inter-correlated items across multiple baskets, necessitating novel theoretical frameworks. The emerging significance of Next-Basket Recommendation (NBR) underscores its importance as a research area. NBRs leverage the user's purchasing history, analyzing sequential series of recently procured baskets to anticipate the next assortment of items. This approach models both enduring and ephemeral preferences, enhancing precision in the recommendation process.

Within the realm of user transaction records, typically manifested as a sequence of shopping baskets, the objective of Next-Basket Recommendation resides in prognosticating the subsequent basket of items a user is inclined to purchase. This is achieved through the modeling of sequential dependencies inherent in the sequence of baskets. A multitude of pertinent studies in the domain of next-basket recommendation has yielded noteworthy achievements. One such noteworthy endeavor, articulated by Rendle et al. \cite{r2}, entails a Markov Chain (MC)-based approach, adept at capturing low-order dependencies across baskets to enhance next-basket recommendation. Additionally, embedding-based methodologies elucidated by Wan et al. \cite{r5} and Wang et al. \cite{r15}, employ distributed representations for predicting the contents of the subsequent basket. Meanwhile, recurrent neural network (RNN) based approaches \cite{r1,r12,r13,r16}, are designed to encapsulate higher-order dependencies spanning multiple baskets. Furthermore, intention-driven strategies, as proposed by Wang et al. \cite{wang2020intention, wang2021intention2basket}, seek to model the heterogeneous intentions inherent in historical sequences of purchased baskets, thereby tailoring next-basket recommendations to align with diverse user intentions. In an alternative domain, a method employing a K-nearest neighbor (KNN) framework \cite{hu2020modeling}, aimed to leverage personalized item frequency information for enhancing the efficacy of NBRs. Similarly, an alternative strategy \cite{r14}, employed a combination of KNN and collaborative filtering (CF) to address the temporal dimension, specifically focusing on modeling the recency of items within the context of NBRs.

Despite notable successes in various studies, a discernible gap exists in articulating the latest advancements and objectively assessing the efficacy of diverse NBR methodologies. The lack of standardized problem formalization and experimental settings further complicates equitable comparisons, exacerbated by inconsistencies in dataset selection and baselines. This hampers the clarity and comparability of research outcomes in the NBR field. This research aims to address these gaps by defining the NBR problem, summarizing existing research, categorizing and comparing NBR approaches, establishing a standardized experimental setup, and conducting a systematic empirical study on selected methodologies. The study concludes by outlining directions for future NBR research, providing valuable insights for researchers in the field.

The main contributions of this work are summarized below: 
\begin{itemize}
\item We present a novel taxonomy to categorize and organize diverse NBR approaches, offering a concise overview of current research progress in the NBR field.

\item 
We systematically examine representative and state-of-the-art NBR approaches through a comprehensive empirical study, offering a unified evaluation for comparison of their performance. This aims to provide valuable insights for future research in this dynamic field.

\item This work represents the inaugural systematic exploration of NBR approaches, incorporating comprehensive evaluation metrics in the existing literature.


\end{itemize}

\section{RELATED WORK}



\subsection{Related Surveys}


Numerous studies have explored next-basket recommendation, encompassing session-based and sequential recommendation. Wang et al. conducted a thorough survey \cite{wang2021survey,wang2020hierarchical}, defining research problems, highlighting challenges, and delineating progress and future directions in session-based recommendation. Ludewig et al. \cite{r26} performed an empirical study, comprehensively comparing session-based recommendation algorithms. In the realm of sequential recommendation, Wang et al. \cite{r7} outlined key challenges, progress, and future directions, while Fang et al. \cite{r9} summarized and compared deep learning-based approaches for sequential recommendation.  While session-based recommendation and sequential recommendation share relevance, they diverge significantly from next-basket recommendation. Session-based and sequential recommendation operate on session and sequence data, respectively, where input comprises sets or sequences of items. Notably, there is an absence of a clear basket structure within these data types. They primarily focus on recommending the next item within the same basket by modeling intra-basket dependencies. In contrast, next-basket recommendation utilizes a sequence of baskets as input, aiming to recommend the subsequent basket by modeling inter-basket dependencies.

Extensive reviews and empirical studies have been conducted on session-based recommendation and sequential recommendation areas\cite{guo2021sequential,hu2017diversifying,wang2022exploiting}. However, there is a notable absence of systematic investigations offering a comprehensive overview of studies in the NBR domain, as well as a lack of unified and rigorous evaluations of various NBR algorithms. Given the increasing popularity of NBR and the evolving research landscape, there is an urgent need for a systematic and thorough examination, summarization, and evaluation of NBR approaches. This paper represents the inaugural effort towards addressing this gap, focusing on delineating the problem statement, categorizing existing NBR approaches, providing benchmarking evaluations, and offering insights into future prospects in the field.




\subsection{Next Basket Recommendation vs. Bundle Recommendation}

Numerous studies explore tasks related to NBR, with bundle recommendation being a prominent example \cite{r24}. Despite apparent similarities, distinctions in settings and assumptions exist between next-basket recommendation and bundle recommendation. The propensity for researchers to conflate these concepts underscores the importance of a comparative analysis to delineate the unique characteristics of next-basket recommendation as distinct from bundle recommendation.
Next basket recommendation and bundle recommendation are contingent upon distinct datasets: basket data and bundle data. To establish a clear understanding, it is imperative to delineate the disparity between these datasets. A basket denotes a collection of items acquired by a user during a single shopping visit. Typically, there is no discernible order among the items within a basket, and their correlation or lack thereof depends on specific circumstances. In sequences of baskets, sorting commonly occurs in ascending order with respect to shopping visits. 
A bundle refers to the amalgamation of two or more closely related items \cite{r25}. These items, within a bundle, are typically unordered and exhibit similarities or complementarity \cite{hao2020p}. When conceptualizing bundles as a sequence, bundle recommendations commonly involve organizing them in descending order based on their respective prices.



Recommendation systems often address the task of suggesting the next set of items for a user by framing it as a sequential prediction problem. This involves predicting a user's future purchases based on their historical transaction data, leveraging the modeled sequential relationships among their past baskets. For instance, if a user buys a printer, recommendations for complementary items like printing paper and toners can be generated for their subsequent basket. In contrast, bundle recommendation, exemplified by \cite{chen2019matching}, is posed as an optimization problem. It seeks to choose an optimal and correlated set of items from a diverse pool to meet a specific consumption goal. Unlike sequential prediction, bundle recommendation does not explicitly consider or model sequential dependencies. For example, a user with a preference for Apple products might receive a bundle recommendation for iPhones and AirPods, disregarding any sequential purchase patterns.

\section{PROBLEM STATEMENT}

\begin{figure}[b]
	\begin{centering}
		\includegraphics[width=0.47\textwidth]{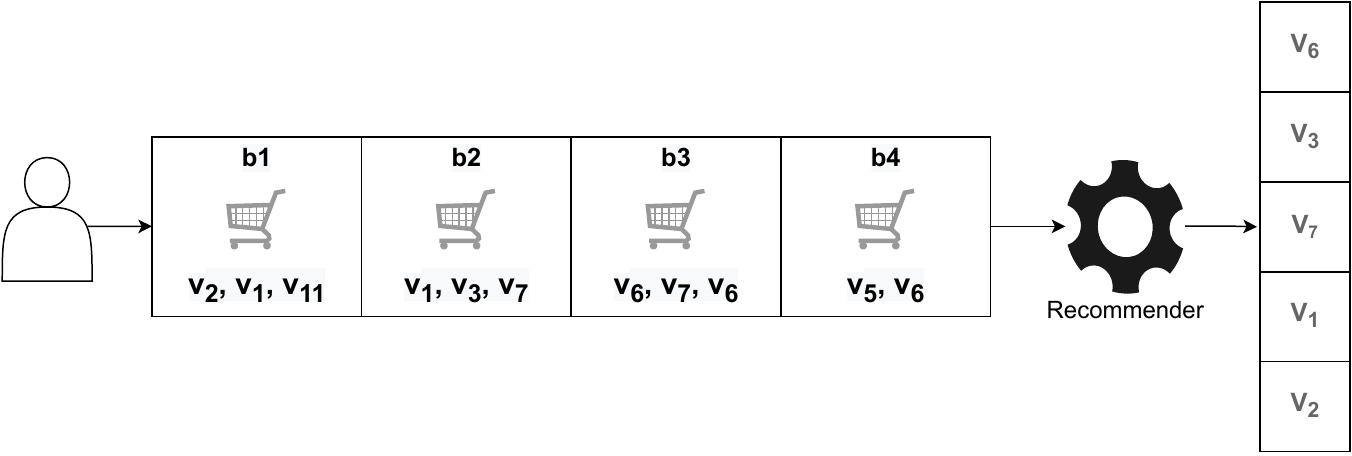}
		\caption{A running example for next basket recommendation.}
		\label{fig:NBRfigure}
	\end{centering}
\end{figure}

Diverse definitions of Next-Basket Recommendation (NBR) tasks emerge across various domains within the literature. For instance, in the realm of E-commerce, a basket is construed not only as a compilation of products procured in a single transaction but also as a collection of places visited during a trip \cite{r13, r19} within the tourism domain. To streamline terminology, we categorize both products and places as "items" in this study. Consequently, we define a "basket" as an assemblage of items collectively purchased in a single transaction event by a specific user. In the context of the next-basket recommendation task, the objective typically involves constructing and training a recommender system based on a user's recently acquired sequence of historical baskets. This system aims to predict the subsequent basket of items that are likely to pique the user's interest by modeling inter-basket dependencies. Typically, this prediction takes the form of generating a personalized, ranked list of items (refer to Figure \ref{fig:NBRfigure}), representing the anticipated contents of the next basket for the given user.




The research problem addressed in this study, as delineated in \cite{wang2020intention}, pertains to the formal definition of the next-basket recommendation task. The dataset under consideration, denoted as $D={s_{1},...,s_{|D|}}$ with $|D|$ representing the total number of sequences, comprises sequences of shopping baskets, hereafter referred to as "baskets." Each sequence $s={b_{1}, ..., b_{|s|}} (s \in D)$ within $D$ represents the historical purchasing behavior of a specific user, arranged chronologically. In this context, a basket $b={v_{1}, ..., v_{|b|}} (b \in s)$ contains items purchased in a single transaction event. The amalgamation of all items across the entire dataset forms the universal item set $V={v_{1},...,v_{|V|}}$, while the universal user set is denoted as $U={u_{1},...,u_{|U|}}$. Figure \ref{fig:NBRfigure} provides an illustrative example for next basket recommendation. Given a sequence of baskets $s={b_{1}, ..., b_t}$, the target basket for prediction is typically the last one, $b_t$. The preceding baskets, denoted as the context $C_t = {b_{1}, ..., b_{t-1}}$, serve as the historical information used in the prediction and recommendation stage. In the next-basket recommendation task, the objective is to predict the items most likely to appear in the user's next basket, denoted as $b_{t}$, based on the context $C_t$ and the user's purchasing history. Formally, this task is expressed as:

\begin{equation}
b_t = f(C_t, u).
\end{equation}

\section{Classes and Comparison of NBR Approaches}
\begin{figure}
    \begin{centering}
    \includegraphics[width=0.50\textwidth]{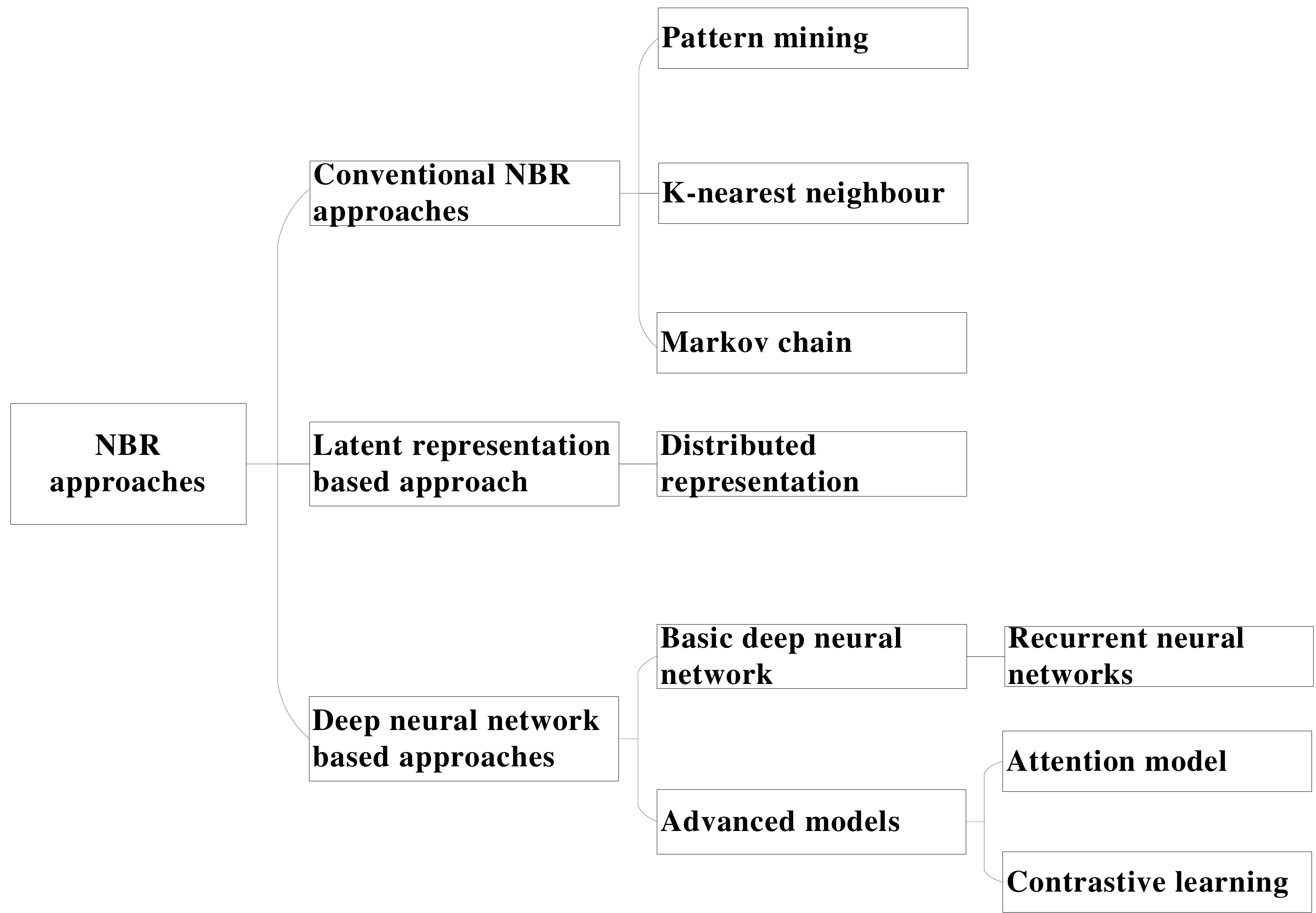}
    \caption{The classes of NBR approaches.}
    \label{fig:categorizationofNBR}
    \end{centering}

\end{figure}

\subsection{Categorization of NBR Approaches.}

The literature delineates three primary classes of NBR approaches, as illustrated in Figure \ref{fig:categorizationofNBR}. These classes are: (1) Conventional NBR approaches, grounded in traditional data mining techniques like pattern mining and K-nearest neighbor (KNN); (2) Latent representation approaches, primarily utilizing representation learning techniques, such as latent embedding; and (3) Deep neural network approaches, typically constructed on deep learning models like recurrent neural networks (RNN). Further granularity is achieved through subdivision, yielding four sub-classes for conventional NBR (pattern mining, KNN, Markov chain-based approaches), one sub-class for latent representation (distributed representation), and two sub-classes for deep neural network approaches (basic and advanced models). The basic deep neural network sub-class encompasses recurrent neural network-based NBR, while the advanced models sub-class further divides into attention models and contrastive learning. Notably, hybrid approaches amalgamate multiple algorithms/models; for instance, ANAM \cite{r16} integrates recurrent neural networks and attention models, and CLEA \cite{qin2021world} combines a contrastive learning framework with a recurrent neural network.

\subsection{A Comparison of Different Classes of NBR Approaches}

Traditional NBR approaches, such as those employing pattern mining \cite{r10}, KNN models \cite{hu2020modeling}, and Markov chain models \cite{r2}, are generally straightforward and efficient due to their reliance on conventional models and algorithms. These methods are particularly suited for simple datasets with less complex dependencies within or between baskets. Despite their simplicity, some, like TIFUKNN \cite{hu2020modeling}, a KNN-based approach, can outperform deep learning-based methods such as Dream and Beacon across various metrics on experimental datasets (cf. TABLE~\ref{tab:statistics_datasets}). In contrast, deep neural network-based approaches are inherently sophisticated, employing diverse neural architectures to capture intrinsic features and dependencies within and between baskets. While these approaches may be more time-consuming and less stable in certain cases, they excel at capturing complex intra- and inter-basket dependencies, leading to superior recommendation performance. For instance, recurrent neural networks (RNNs) have proven effective in handling basket sequences in recent years \cite{r1,r16,r12,r13}. Latent representation-based approaches, in contrast, tend to be more efficient and simpler than deep learning-based methods, as they do not employ complex neural architectures. However, due to the power of embedding techniques, they can effectively capture implicit and hidden patterns in data, resulting in enhanced performance. An illustrative example is HRM \cite{r15}, a distributed representation-based NBR approach utilizing a three-layer structure to construct a user's last basket representation for predicting their next basket.

\section{Datasets and Compared Approaches}

\subsection{Datasets and Data Pre-processing}

Upon analyzing the datasets in the collected papers, two significant issues have been identified. Firstly, certain datasets are not publicly available due to user privacy limitations. Secondly, there are instances where datasets with the same name exist in different versions. Considering the popularity and characteristics of these datasets, we have chosen three commonly utilized datasets for our experiments in evaluating next-basket prediction performance. They are \textbf{TaFeng}\footnote{https://www.kaggle.com/chiranjivdas09/ta-feng-grocery-dataset}, \textbf{Instacart}\footnote{https://www.kaggle.com/c/instacart-market-basket-analysis/data} and \textbf{Dunnhumby}\footnote{https://www.dunnhumby.com/source-files/}. For data pre-processing, the removal of infrequent items and inactive users with minimal interactions constitutes a crucial step in this process. An examination of pertinent literature reveals a lack of uniformity in the pre-processing strategies employed by various studies. To ensure equitable performance assessment across diverse datasets, we uniformly pre-process by filtering out items purchased less than $n$ times, where $n$ assumes values of 10, 20, and 17 for the TaFeng, Instacart, and Dunnhumby datasets, respectively.
\begin{table}[htbp]
 \centering
 \caption{The statistics of experimental datasets.}
 \label{tab:statistics_datasets}
  \begin{tabular}{llll}
  \toprule
  \textbf{Statistics} & \textbf{TaFeng} & \textbf{Instacart} & \textbf{Dunnhumby}\\
  \midrule
  \#Users & 20,212 & 19,982 & 22,530\\
  \#Baskets & 105,140 & 280,941 & 214,861\\
  \#Items & 10,411 & 13,400 & 3,920\\
  \#Basket/user & 5.20 & 14.06 & 9.53\\
  \#Items/basket & 6.32 & 9.61 & 7.45\\
  \bottomrule
 \end{tabular}
 \begin{flushleft}
 \footnotesize
  The rows \#Users, \#Baskets, \#Items, \#Basket/user, \#Items/basket correspond to the number of users, the number of baskets over all users, the number of items, the average number of baskets per user and the average number of  items per basket, respectively.
  \end{flushleft}
\end{table}

\subsection{Compared Approaches and Evaluation Metrics}

Within our paper compilation, diverse approaches are employed to ensure both reproducibility and popularity. Eight selected approaches for experimentation include \textbf{TOP}, \textbf{FPMC} \cite{r2}, \textbf{DREAM} \cite{r1}, \textbf{Beacon} \cite{r13}, \textbf{Sets2Sets} \cite{r12}, \textbf{UP-CF@r} \cite{r14}, \textbf{TIFUKNN} \cite{hu2020modeling}, and \textbf{CLEA} \cite{qin2021world}. For instance, Sets2Sets is tailored for sequential set prediction using an encoder-decoder framework, specifically adapted for NBR.  In addition, to holistically assess various approaches, we employ seven metrics: \textbf{Recall}, \textbf{Precision}, \textbf{F1-Score}, Person-wise Hit Ratio (\textbf{PHR}), Normalized Discounted Cumulative Gain (\textbf{NDCG}), Mean Average Precision (\textbf{MAP}), and Mean Reciprocal Rank (\textbf{MRR}). It is noteworthy that the latter two metrics, namely \textbf{MAP} and \textbf{MRR}, are prevalent in top-$K$ recommendation despite their absence in the collected papers. 
Adhering to the conventions established in traditional NBR work \cite{hu2020modeling,qin2021world}, each approach is evaluated by recommending a predicted basket (i.e., recommendation list) of a predetermined size $K$. Metrics are uniformly calculated across all predicted baskets, with a common attribute: superior performance is indicated by larger values.

\begin{table*}
\caption{The statistics of compared approaches.}
\label{table: statisticsofallmehtods}
\begin{center}
\setlength{\tabcolsep}{4.0mm
\begin{tabular}{cccccccccc}
\toprule[2pt]
\multicolumn{3}{l}{\multirow{1}{*}{Dataset}} &
 \multicolumn{5}{c}{\multirow{1}{*}{TaFeng}}& \\ \hline
\multicolumn{1}{c}{$K$} &
  \multicolumn{2}{l}{Approaches} &
  \multicolumn{1}{c}{Recall} &
  \multicolumn{1}{c}{Precision} &
  \multicolumn{1}{c}{F1-score} &
  \multicolumn{1}{c}{PHR} &
  \multicolumn{1}{c}{MAP} &
  \multicolumn{1}{c}{MRR} &
  \multicolumn{1}{c}{NDCG} \\ \hline
\multicolumn{1}{c}{\multirow{8}{*}{5}} &
  \multicolumn{2}{l}{TOP} &
  \multicolumn{1}{c}{0.0670} &
  \multicolumn{1}{c}{0.0420} &
  \multicolumn{1}{c}{0.0412} &
  \multicolumn{1}{c}{0.1870} &
  \multicolumn{1}{c}{0.1426} &
  \multicolumn{1}{c}{0.1438} &
  \multicolumn{1}{c}{0.0733} \\ \cline{2-10} 
\multicolumn{1}{c}{} &
  \multicolumn{2}{l}{FPMC} &
  \multicolumn{1}{c}{0.0632} &
  \multicolumn{1}{c}{0.0433} &
  \multicolumn{1}{c}{0.0514} &
  \multicolumn{1}{c}{0.1912} &
  \multicolumn{1}{c}{0.1066} &
  \multicolumn{1}{c}{0.1070} &
  \multicolumn{1}{c}{0.0557} \\ \cline{2-10} 
\multicolumn{1}{c}{} &
  \multicolumn{2}{l}{DREAM} &
  \multicolumn{1}{c}{0.0921} &
  \multicolumn{1}{c}{0.0533} &
  \multicolumn{1}{c}{0.0675} &
  \multicolumn{1}{c}{0.2329} &
  \multicolumn{1}{c}{0.1580} &
  \multicolumn{1}{c}{0.1608} &
  \multicolumn{1}{c}{0.0891} \\ \cline{2-10} 
\multicolumn{1}{c}{} &
  \multicolumn{2}{l}{Beacon} &
  \multicolumn{1}{c}{0.0820} &
  \multicolumn{1}{c}{0.0489} &
  \multicolumn{1}{c}{0.0493} &
  \multicolumn{1}{c}{0.2117} &
  \multicolumn{1}{c}{0.1473} &
  \multicolumn{1}{c}{0.1510} &
  \multicolumn{1}{c}{0.0817} \\ \cline{2-10}
\multicolumn{1}{c}{} &
  \multicolumn{2}{l}{Sets2Sets} &
  \multicolumn{1}{c}{0.0951} &
  \multicolumn{1}{c}{0.0690} &
  \multicolumn{1}{c}{0.0647} &
  \multicolumn{1}{c}{0.2860} &
  \multicolumn{1}{c}{0.1718} &
  \multicolumn{1}{c}{0.1763} &
  \multicolumn{1}{c}{0.0925} \\ \cline{2-10}
\multicolumn{1}{c}{} &
  \multicolumn{2}{l}{UP-CF@r} &
  \multicolumn{1}{c}{0.0741} &
  \multicolumn{1}{c}{0.0644} &
  \multicolumn{1}{c}{0.0565} &
  \multicolumn{1}{c}{0.2601} &
  \multicolumn{1}{c}{0.1539} &
  \multicolumn{1}{c}{0.1574} &
  \multicolumn{1}{c}{0.0726} \\ \cline{2-10} 
\multicolumn{1}{c}{} &
   \multicolumn{2}{l}{TIFUKNN} &
  \multicolumn{1}{c}{0.0833} &
  \multicolumn{1}{c}{0.0713} &
  \multicolumn{1}{c}{0.0624} &
  \multicolumn{1}{c}{0.2799} &
  \multicolumn{1}{c}{0.1626} &
  \multicolumn{1}{c}{0.1663} &
  \multicolumn{1}{c}{0.0790} \\ \cline{2-10}
\multicolumn{1}{c}{} &
  \multicolumn{2}{l}{CLEA} &
  \multicolumn{1}{c}{\textbf{0.1280}} &
  \multicolumn{1}{c}{\textbf{0.0900}} &
  \multicolumn{1}{c}{\textbf{0.0842}} &
  \multicolumn{1}{c}{\textbf{0.3547}} &
  \multicolumn{1}{c}{\textbf{0.2321}} &
  \multicolumn{1}{c}{\textbf{0.2386}} &
  \multicolumn{1}{c}{\textbf{0.1284}} \\ 
  \hline
\multicolumn{1}{c}{\multirow{7}{*}{10}} &
  \multicolumn{2}{l}{TOP} &
  \multicolumn{1}{c}{0.0768} &
  \multicolumn{1}{c}{0.0269} &
  \multicolumn{1}{c}{0.0326} &
  \multicolumn{1}{c}{0.2282} &
  \multicolumn{1}{c}{0.1455} &
  \multicolumn{1}{c}{0.1495} &
  \multicolumn{1}{c}{0.0786} \\ \cline{2-10} 
\multicolumn{1}{c}{} &
  \multicolumn{2}{l}{FPMC} &
  \multicolumn{1}{c}{0.0755} &
  \multicolumn{1}{c}{0.0274} &
  \multicolumn{1}{c}{0.0402} &
  \multicolumn{1}{c}{0.2314} &
  \multicolumn{1}{c}{0.1096} &
  \multicolumn{1}{c}{0.1122} &
  \multicolumn{1}{c}{0.0616} \\ \cline{2-10} 
\multicolumn{1}{c}{} &
  \multicolumn{2}{l}{DREAM} &
  \multicolumn{1}{c}{0.1170} &
  \multicolumn{1}{c}{0.0362} &
  \multicolumn{1}{c}{0.0553} &
  \multicolumn{1}{c}{0.3060} &
  \multicolumn{1}{c}{0.1648} &
  \multicolumn{1}{c}{0.1705} &
  \multicolumn{1}{c}{0.1004} \\ \cline{2-10} 
\multicolumn{1}{c}{} &
  \multicolumn{2}{l}{Beacon} &
  \multicolumn{1}{c}{0.1059} &
  \multicolumn{1}{c}{0.0348} &
  \multicolumn{1}{c}{0.0435} &
  \multicolumn{1}{c}{0.2863} &
  \multicolumn{1}{c}{0.1543} &
  \multicolumn{1}{c}{0.1608} &
  \multicolumn{1}{c}{0.0930} \\ \cline{2-10} 
\multicolumn{1}{c}{} &
  \multicolumn{2}{l}{Sets2Sets} &
  \multicolumn{1}{c}{0.1315} &
  \multicolumn{1}{c}{0.0512} &
  \multicolumn{1}{c}{0.0614} &
  \multicolumn{1}{c}{0.3797} &
  \multicolumn{1}{c}{0.1730} &
  \multicolumn{1}{c}{0.1816} &
  \multicolumn{1}{c}{0.1082} \\ \cline{2-10} 
\multicolumn{1}{c}{} &
  \multicolumn{2}{l}{UP-CF@r} &
  \multicolumn{1}{c}{0.1058} &
  \multicolumn{1}{c}{0.0496} &
  \multicolumn{1}{c}{0.0560} &
  \multicolumn{1}{c}{0.3483} &
  \multicolumn{1}{c}{0.1600} &
  \multicolumn{1}{c}{0.1691} &
  \multicolumn{1}{c}{0.0889} \\ \cline{2-10} 
\multicolumn{1}{c}{} &
  \multicolumn{2}{l}{TIFUKNN} &
  \multicolumn{1}{c}{0.1236} &
  \multicolumn{1}{c}{0.0537} &
  \multicolumn{1}{c}{0.0619} &
  \multicolumn{1}{c}{0.3765} &
  \multicolumn{1}{c}{0.1680} &
  \multicolumn{1}{c}{0.1791} &
  \multicolumn{1}{c}{0.0982} \\
  \cline{2-10}
\multicolumn{1}{c}{} &
  \multicolumn{2}{l}{CLEA} &
  \multicolumn{1}{c}{\textbf{0.1609}} &
  \multicolumn{1}{c}{\textbf{0.0648}} &
  \multicolumn{1}{c}{\textbf{0.0760}} &
  \multicolumn{1}{c}{\textbf{0.4446}} &
  \multicolumn{1}{c}{\textbf{0.2353}} &
  \multicolumn{1}{c}{\textbf{0.2514}} &
  \multicolumn{1}{c}{\textbf{0.1472}}\\ 
 \toprule[2pt]
\multicolumn{3}{l}{\multirow{1}{*}{Dataset}} &
\multicolumn{5}{c}{\multirow{1}{*}{Instacart}}& \\ \hline
\multicolumn{1}{c}{$K$} &
  \multicolumn{2}{l}{Approaches} &
  \multicolumn{1}{c}{Recall} &
  \multicolumn{1}{c}{Precision} &
  \multicolumn{1}{c}{F1-score} &
  \multicolumn{1}{c}{PHR} &
  \multicolumn{1}{c}{MAP} &
  \multicolumn{1}{c}{MRR} &
  \multicolumn{1}{c}{NDCG} \\ \hline
\multicolumn{1}{c}{\multirow{8}{*}{5}} &
  \multicolumn{2}{l}{TOP} &
  \multicolumn{1}{c}{0.0487} &
  \multicolumn{1}{c}{0.0956} &
  \multicolumn{1}{c}{0.0581} &
  \multicolumn{1}{c}{0.3668} &
  \multicolumn{1}{c}{0.2270} &
  \multicolumn{1}{c}{0.2321} &
  \multicolumn{1}{c}{0.0666} \\ \cline{2-10} 
\multicolumn{1}{c}{} &
  \multicolumn{2}{l}{FPMC} &
  \multicolumn{1}{c}{0.0481} &
  \multicolumn{1}{c}{0.0948} &
  \multicolumn{1}{c}{0.0620} &
  \multicolumn{1}{c}{0.3600} &
  \multicolumn{1}{c}{0.2211} &
  \multicolumn{1}{c}{0.2264} &
  \multicolumn{1}{c}{0.0651} \\ \cline{2-10} 
\multicolumn{1}{c}{} &
  \multicolumn{2}{l}{DREAM} &
  \multicolumn{1}{c}{0.0763} &
  \multicolumn{1}{c}{0.0486} &
  \multicolumn{1}{c}{0.0594} &
  \multicolumn{1}{c}{0.2176} &
  \multicolumn{1}{c}{0.1543} &
  \multicolumn{1}{c}{0.1481} &
  \multicolumn{1}{c}{0.0754} \\ \cline{2-10} 
\multicolumn{1}{c}{} &
  \multicolumn{2}{l}{Beacon} &
  \multicolumn{1}{c}{0.0539} &
  \multicolumn{1}{c}{0.1049} &
  \multicolumn{1}{c}{0.0638} &
  \multicolumn{1}{c}{0.3930} &
  \multicolumn{1}{c}{0.2181} &
  \multicolumn{1}{c}{0.2250} &
  \multicolumn{1}{c}{0.0695} \\ \cline{2-10}
\multicolumn{1}{c}{} &
  \multicolumn{2}{l}{Sets2Sets} &
  \multicolumn{1}{c}{0.1266} &
  \multicolumn{1}{c}{0.1652} &
  \multicolumn{1}{c}{0.1209} &
  \multicolumn{1}{c}{0.5503} &
  \multicolumn{1}{c}{0.3141} &
  \multicolumn{1}{c}{0.3260} &
  \multicolumn{1}{c}{0.1375} \\ \cline{2-10}
\multicolumn{1}{c}{} &
  \multicolumn{2}{l}{UP-CF@r} &
  \multicolumn{1}{c}{0.2512} &
  \multicolumn{1}{c}{0.3580} &
  \multicolumn{1}{c}{0.2512} &
  \multicolumn{1}{c}{0.7946} &
  \multicolumn{1}{c}{0.5818} &
  \multicolumn{1}{c}{0.6138} &
  \multicolumn{1}{c}{0.2970} \\ \cline{2-10} 
\multicolumn{1}{c}{} &
   \multicolumn{2}{l}{TIFUKNN} &
  \multicolumn{1}{c}{\textbf{0.2616}} &
  \multicolumn{1}{c}{\textbf{0.3733}} &
  \multicolumn{1}{c}{\textbf{0.2619}} &
  \multicolumn{1}{c}{\textbf{0.8052}} &
  \multicolumn{1}{c}{\textbf{0.5992}} &
  \multicolumn{1}{c}{\textbf{0.6312}} &
  \multicolumn{1}{c}{\textbf{0.3094}} \\ \cline{2-10}
\multicolumn{1}{c}{} &
  \multicolumn{2}{l}{CLEA} &
  \multicolumn{1}{c}{0.1850} &
  \multicolumn{1}{c}{0.3025} &
  \multicolumn{1}{c}{0.1973} &
  \multicolumn{1}{c}{0.7136} &
  \multicolumn{1}{c}{0.5623} &
  \multicolumn{1}{c}{0.5865} &
  \multicolumn{1}{c}{0.2450} \\ 
  \hline
\multicolumn{1}{c}{\multirow{7}{*}{10}} &
  \multicolumn{2}{l}{TOP} &
  \multicolumn{1}{c}{0.0731} &
  \multicolumn{1}{c}{0.0735} &
  \multicolumn{1}{c}{0.0655} &
  \multicolumn{1}{c}{0.4614} &
  \multicolumn{1}{c}{0.2223} &
  \multicolumn{1}{c}{0.2448} &
  \multicolumn{1}{c}{0.0832} \\ \cline{2-10} 
\multicolumn{1}{c}{} &
  \multicolumn{2}{l}{FPMC} &
  \multicolumn{1}{c}{0.0712} &
  \multicolumn{1}{c}{0.0716} &
  \multicolumn{1}{c}{0.0697} &
  \multicolumn{1}{c}{0.4508} &
  \multicolumn{1}{c}{0.2182} &
  \multicolumn{1}{c}{0.2388} &
  \multicolumn{1}{c}{0.0809} \\ \cline{2-10} 
\multicolumn{1}{c}{} &
  \multicolumn{2}{l}{DREAM} &
  \multicolumn{1}{c}{0.1012} &
  \multicolumn{1}{c}{0.0359} &
  \multicolumn{1}{c}{0.0530} &
  \multicolumn{1}{c}{0.3037} &
  \multicolumn{1}{c}{0.1543} &
  \multicolumn{1}{c}{0.1596} &
  \multicolumn{1}{c}{0.0874} \\ \cline{2-10} 
\multicolumn{1}{c}{} &
  \multicolumn{2}{l}{Beacon} &
  \multicolumn{1}{c}{0.0767} &
  \multicolumn{1}{c}{0.0769} &
  \multicolumn{1}{c}{0.0686} &
  \multicolumn{1}{c}{0.4755} &
  \multicolumn{1}{c}{0.2174} &
  \multicolumn{1}{c}{0.2362} &
  \multicolumn{1}{c}{0.0851} \\ \cline{2-10}
\multicolumn{1}{c}{} &
  \multicolumn{2}{l}{Sets2Sets} &
  \multicolumn{1}{c}{0.2058} &
  \multicolumn{1}{c}{0.1463} &
  \multicolumn{1}{c}{0.1710} &
  \multicolumn{1}{c}{0.7157} &
  \multicolumn{1}{c}{0.3155} &
  \multicolumn{1}{c}{0.3550} &
  \multicolumn{1}{c}{0.1862} \\ \cline{2-10}
\multicolumn{1}{c}{} &
  \multicolumn{2}{l}{UP-CF@r} &
  \multicolumn{1}{c}{0.3480} &
  \multicolumn{1}{c}{0.2731} &
  \multicolumn{1}{c}{0.2650} &
  \multicolumn{1}{c}{0.8590} &
  \multicolumn{1}{c}{0.5448} &
  \multicolumn{1}{c}{0.6226} &
  \multicolumn{1}{c}{0.3613} \\ \cline{2-10} 
\multicolumn{1}{c}{} &
   \multicolumn{2}{l}{TIFUKNN} &
  \multicolumn{1}{c}{\textbf{0.3698}} &
  \multicolumn{1}{c}{\textbf{0.2896}} &
  \multicolumn{1}{c}{\textbf{0.2812}} &
  \multicolumn{1}{c}{\textbf{0.8756}} &
  \multicolumn{1}{c}{\textbf{0.5618}} &
  \multicolumn{1}{c}{\textbf{0.6409}} &
  \multicolumn{1}{c}{\textbf{0.3805}} \\ \cline{2-10}
\multicolumn{1}{c}{} &
  \multicolumn{2}{l}{CLEA} &
  \multicolumn{1}{c}{0.2135} &
  \multicolumn{1}{c}{0.1857} &
  \multicolumn{1}{c}{0.1738} &
  \multicolumn{1}{c}{0.7439} &
  \multicolumn{1}{c}{0.5020} &
  \multicolumn{1}{c}{0.5659} &
  \multicolumn{1}{c}{0.2537} \\ 
 \toprule[2pt]
\multicolumn{3}{l}{\multirow{1}{*}{Dataset}} &
\multicolumn{5}{c}{\multirow{1}{*}{Dunnhumby}}& \\ \hline
\multicolumn{1}{c}{$K$} &
  \multicolumn{2}{l}{Approaches} &
  \multicolumn{1}{c}{Recall} &
  \multicolumn{1}{c}{Precision} &
  \multicolumn{1}{c}{F1-score} &
  \multicolumn{1}{c}{PHR} &
  \multicolumn{1}{c}{MAP} &
  \multicolumn{1}{c}{MRR} &
  \multicolumn{1}{c}{NDCG} \\ \hline
\multicolumn{1}{c}{\multirow{8}{*}{5}} &
  \multicolumn{2}{l}{TOP} &
  \multicolumn{1}{c}{0.0966} &
  \multicolumn{1}{c}{0.0763} &
  \multicolumn{1}{c}{0.0606} &
  \multicolumn{1}{c}{0.3294} &
  \multicolumn{1}{c}{0.2190} &
  \multicolumn{1}{c}{0.2282} &
  \multicolumn{1}{c}{0.0861} \\ \cline{2-10} 
\multicolumn{1}{c}{} &
  \multicolumn{2}{l}{FPMC} &
  \multicolumn{1}{c}{0.0746} &
  \multicolumn{1}{c}{0.1033} &
  \multicolumn{1}{c}{0.0866} &
  \multicolumn{1}{c}{0.3928} &
  \multicolumn{1}{c}{0.2566} &
  \multicolumn{1}{c}{0.2674} &
  \multicolumn{1}{c}{0.0903} \\ \cline{2-10} 
\multicolumn{1}{c}{} &
  \multicolumn{2}{l}{DREAM} &
  \multicolumn{1}{c}{0.0756} &
  \multicolumn{1}{c}{0.1049} &
  \multicolumn{1}{c}{0.0879} &
  \multicolumn{1}{c}{0.3984} &
  \multicolumn{1}{c}{0.2616} &
  \multicolumn{1}{c}{0.2726} &
  \multicolumn{1}{c}{0.0918} \\ \cline{2-10} 
\multicolumn{1}{c}{} &
  \multicolumn{2}{l}{Beacon} &
  \multicolumn{1}{c}{0.0795} &
  \multicolumn{1}{c}{0.1038} &
  \multicolumn{1}{c}{0.0737} &
  \multicolumn{1}{c}{0.3948} &
  \multicolumn{1}{c}{0.2517} &
  \multicolumn{1}{c}{0.2682} &
  \multicolumn{1}{c}{0.0932} \\ \cline{2-10}
\multicolumn{1}{c}{} &
  \multicolumn{2}{l}{Sets2Sets} &
  \multicolumn{1}{c}{0.1040} &
  \multicolumn{1}{c}{0.1249} &
  \multicolumn{1}{c}{0.0904} &
  \multicolumn{1}{c}{0.4353} &
  \multicolumn{1}{c}{0.2515} &
  \multicolumn{1}{c}{0.2867} &
  \multicolumn{1}{c}{0.1112} \\ \cline{2-10}
\multicolumn{1}{c}{} &
  \multicolumn{2}{l}{UP-CF@r} &
  \multicolumn{1}{c}{\textbf{0.1794}} &
  \multicolumn{1}{c}{\textbf{0.2417}} &
  \multicolumn{1}{c}{\textbf{0.1693}} &
  \multicolumn{1}{c}{\textbf{0.6005}} &
  \multicolumn{1}{c}{\textbf{0.4367}} &
  \multicolumn{1}{c}{\textbf{0.4602}} &
  \multicolumn{1}{c}{\textbf{0.2105}} \\ \cline{2-10} 
\multicolumn{1}{c}{} &
   \multicolumn{2}{l}{TIFUKNN} &
  \multicolumn{1}{c}{0.1725} &
  \multicolumn{1}{c}{0.2329} &
  \multicolumn{1}{c}{0.1630} &
  \multicolumn{1}{c}{0.5947} &
  \multicolumn{1}{c}{0.4260} &
  \multicolumn{1}{c}{0.4465} &
  \multicolumn{1}{c}{0.2027} \\ \cline{2-10}
\multicolumn{1}{c}{} &
  \multicolumn{2}{l}{CLEA} &
  \multicolumn{1}{c}{0.1193} &
  \multicolumn{1}{c}{0.1720} &
  \multicolumn{1}{c}{0.1165} &
  \multicolumn{1}{c}{0.5042} &
  \multicolumn{1}{c}{0.3512} &
  \multicolumn{1}{c}{0.3663} &
  \multicolumn{1}{c}{0.1422} \\ 
  \hline
\multicolumn{1}{c}{\multirow{7}{*}{10}} &
  \multicolumn{2}{l}{TOP} &
  \multicolumn{1}{c}{0.1169} &
  \multicolumn{1}{c}{0.0566} &
  \multicolumn{1}{c}{0.0555} &
  \multicolumn{1}{c}{0.4161} &
  \multicolumn{1}{c}{0.2194} &
  \multicolumn{1}{c}{0.2385} &
  \multicolumn{1}{c}{0.0983} \\ \cline{2-10} 
\multicolumn{1}{c}{} &
  \multicolumn{2}{l}{FPMC} &
  \multicolumn{1}{c}{0.0961} &
  \multicolumn{1}{c}{0.0705} &
  \multicolumn{1}{c}{0.0813} &
  \multicolumn{1}{c}{0.4560} &
  \multicolumn{1}{c}{0.2483} &
  \multicolumn{1}{c}{0.2758} &
  \multicolumn{1}{c}{0.1037} \\ \cline{2-10} 
\multicolumn{1}{c}{} &
  \multicolumn{2}{l}{DREAM} &
  \multicolumn{1}{c}{0.1006} &
  \multicolumn{1}{c}{0.0732} &
  \multicolumn{1}{c}{0.0847} &
  \multicolumn{1}{c}{0.4711} &
  \multicolumn{1}{c}{0.2535} &
  \multicolumn{1}{c}{0.2822} &
  \multicolumn{1}{c}{0.1069} \\ \cline{2-10} 
\multicolumn{1}{c}{} &
  \multicolumn{2}{l}{Beacon} &
  \multicolumn{1}{c}{0.1030} &
  \multicolumn{1}{c}{0.0726} &
  \multicolumn{1}{c}{0.0709} &
  \multicolumn{1}{c}{0.4667} &
  \multicolumn{1}{c}{0.2493} &
  \multicolumn{1}{c}{0.2776} &
  \multicolumn{1}{c}{0.1079} \\ \cline{2-10}
\multicolumn{1}{c}{} &
  \multicolumn{2}{l}{Sets2Sets} &
  \multicolumn{1}{c}{0.1695} &
  \multicolumn{1}{c}{0.1102} &
  \multicolumn{1}{c}{0.1091} &
  \multicolumn{1}{c}{0.5751} &
  \multicolumn{1}{c}{0.2563} &
  \multicolumn{1}{c}{0.2867} &
  \multicolumn{1}{c}{0.1488} \\ \cline{2-10}
\multicolumn{1}{c}{} &
  \multicolumn{2}{l}{UP-CF@r} &
  \multicolumn{1}{c}{\textbf{0.2480}} &
  \multicolumn{1}{c}{\textbf{0.1795}} &
  \multicolumn{1}{c}{\textbf{0.1733}} &
  \multicolumn{1}{c}{\textbf{0.6764}} &
  \multicolumn{1}{c}{\textbf{0.4149}} &
  \multicolumn{1}{c}{\textbf{0.4703}} &
  \multicolumn{1}{c}{\textbf{0.2525}} \\ \cline{2-10} 
\multicolumn{1}{c}{} &
   \multicolumn{2}{l}{TIFUKNN} &
  \multicolumn{1}{c}{0.2419} &
  \multicolumn{1}{c}{0.1734} &
  \multicolumn{1}{c}{0.1677} &
  \multicolumn{1}{c}{0.6713} &
  \multicolumn{1}{c}{0.4054} &
  \multicolumn{1}{c}{0.4568} &
  \multicolumn{1}{c}{0.2444} \\ \cline{2-10}
\multicolumn{1}{c}{} &
  \multicolumn{2}{l}{CLEA} &
  \multicolumn{1}{c}{0.1476} &
  \multicolumn{1}{c}{0.1102} &
  \multicolumn{1}{c}{0.1048} &
  \multicolumn{1}{c}{0.5555} &
  \multicolumn{1}{c}{0.3555} &
  \multicolumn{1}{c}{0.3861} &
  \multicolumn{1}{c}{0.1673} \\ 
  \toprule[2pt]
\end{tabular}
 }
\end{center}
\end{table*}



\section{Performance of Compared Approaches}

\textit{\textbf{Finding 1:}} Conventional approaches, such as TOP, FPMC, UP-CF@r, and TIFUKNN, exhibit remarkable performance. TOP, the simplest approach, often outperforms FPMC in terms of Recall, MAP, MRR, and NDCG across most datasets, emphasizing the significance of item frequency information in next-basket recommendation. This underscores the utility of TOP as a baseline for evaluating new models \cite{ariannezhad2022recanet}. On Instacart and Dunnhumby datasets, UP-CF@r and TIFUKNN consistently demonstrate superior performance across all metrics, leveraging item attributes and a user-based nearest neighbor approach. Despite their simplicity, conventional approaches prove highly effective on suitable datasets.


\textit{\textbf{Finding 2:}} Deep neural network approaches, such as DREAM, Beacon, Sets2Sets, and CLEA, lack consistent and stable performance. Notably, CLEA outperforms others on the TaFeng dataset by automatically capturing interactions between historical items and the target item. However, its superiority does not extend to the Instacart and Dunnhumby datasets, possibly due to dataset characteristics, specifically the lower average basket size in TaFeng. Sets2Sets, utilizing a repeated pattern based on an encoder-decoder architecture, surpasses DREAM, Beacon, and even CLEA in the Dunnhumby dataset, highlighting the prevalence of repeated purchases in the NBR task.
\textit{\textbf{Finding 3:}} 
Conventional approaches generally outperform deep neural network-based methods across most datasets. However, the performance of different approaches varies significantly across datasets, with no single approach excelling on all of them. Analyzing performance on three datasets using F1-Score and NDCG, as presented in Table \ref{table: statisticsofallmehtods}, reveals Sets2Sets, UP-CF@r, TIFUKNN, and CLEA as notable performers. While Sets2Sets and CLEA exhibit superior performance on the Tafeng dataset compared to UP-CF@r and TIFUKNN, they are outperformed on Instacart and Dunnhumby datasets. This discrepancy suggests that Sets2Sets and CLEA may be more suitable for datasets with shorter baskets, whereas UP-CF@r and TIFUKNN excel in recommending next baskets for datasets with longer baskets.

\section{Conclusion}


This paper systematically reviews NBR task literature, delineating distinctions from traditional RSs and articulating the NBR problem statement. We categorize existing work into conventional, latent representation, and deep neural network approaches, analyzing their pros and cons. To enhance understanding and evaluation, we scrutinize popular NBR algorithms and metrics, conducting comparative performance assessments with unified experimental settings. Our unified framework offers a valuable reference for NBR research, aiming to provide clarity and fairness in the evaluation of NBRs for readers and researchers alike.


\bibliographystyle{IEEEtran}
\bibliography{ref}

\begin{thebibliography}{10}
\providecommand{\url}[1]{#1}
\csname url@samestyle\endcsname
\providecommand{\newblock}{\relax}
\providecommand{\bibinfo}[2]{#2}
\providecommand{\BIBentrySTDinterwordspacing}{\spaceskip=0pt\relax}
\providecommand{\BIBentryALTinterwordstretchfactor}{4}
\providecommand{\BIBentryALTinterwordspacing}{\spaceskip=\fontdimen2\font plus
\BIBentryALTinterwordstretchfactor\fontdimen3\font minus \fontdimen4\font\relax}
\providecommand{\BIBforeignlanguage}[2]{{%
\expandafter\ifx\csname l@#1\endcsname\relax
\typeout{** WARNING: IEEEtran.bst: No hyphenation pattern has been}%
\typeout{** loaded for the language `#1'. Using the pattern for}%
\typeout{** the default language instead.}%
\else
\language=\csname l@#1\endcsname
\fi
#2}}
\providecommand{\BIBdecl}{\relax}
\BIBdecl

\bibitem{qin2021world}
Y.~Qin, P.~Wang, and C.~Li, ``The world is binary: Contrastive learning for denoising next basket recommendation,'' in \emph{SIGIR}, 2021, pp. 859--868.

\bibitem{qian2022}
Q.~Zhang, S.~Wang, W.~Lu, C.~Feng, X.~Peng, and Q.~Wang, ``Rethinking adjacent dependency in session-based recommendations,'' in \emph{PAKDD}, 2022, pp. 301--313.

\bibitem{lu2022aspect}
W.~Lu, R.~Wang, S.~Wang, X.~Peng, H.~Wu, and Q.~Zhang, ``Aspect-driven user preference and news representation learning for news recommendation,'' \emph{IEEE Transactions on Intelligent Transportation Systems}, 2022.

\bibitem{lu2019graph}
W.~Lu, F.~Meng, S.~Wang, G.~Zhang, X.~Zhang, A.~Ouyang, and X.~Zhang, ``Graph-based chinese word sense disambiguation with multi-knowledge integration,'' \emph{Computers, Materials \& Continua}, vol.~61, no.~1, pp. 197--212, 2019.

\bibitem{lu2021sentence}
W.~Lu, R.~Yu, S.~Wang, C.~Wang, P.~Jian, and H.~Huang, ``Sentence semantic matching based on 3d cnn for human--robot language interaction,'' \emph{ACM Transactions on Internet Technology}, vol.~21, no.~4, pp. 1--24, 2021.

\bibitem{wang2021survey}
S.~Wang, L.~Cao, Y.~Wang, Q.~Z. Sheng, M.~A. Orgun, and D.~Lian, ``A survey on session-based recommender systems,'' \emph{ACM Computing Surveys}, vol.~54, no.~7, pp. 1--38, 2021.

\bibitem{song2021next}
W.~Song, S.~Wang, Y.~Wang, and S.~Wang, ``Next-item recommendations in short sessions,'' in \emph{RecSys}, 2021, pp. 282--291.

\bibitem{wang2020modelling}
N.~Wang, S.~Wang, Y.~Wang, Q.~Z. Sheng, and M.~Orgun, ``Modelling local and global dependencies for next-item recommendations,'' in \emph{WISE}, 2020, pp. 285--300.

\bibitem{wang2019modeling}
S.~Wang, L.~Hu, Y.~Wang, Q.~Z. Sheng, M.~Orgun, and L.~Cao, ``Modeling multi-purpose sessions for next-item recommendations via mixture-channel purpose routing networks,'' in \emph{IJCAI}, 2019, pp. 3771--3777.

\bibitem{wang2018attention}
S.~Wang, L.~Hu, L.~Cao, X.~Huang, D.~Lian, and W.~Liu, ``Attention-based transactional context embedding for next-item recommendation,'' in \emph{AAAI}, 2018, pp. 2532--2539.

\bibitem{r2}
S.~Rendle, C.~Freudenthaler, and L.~Schmidt-Thieme, ``Factorizing personalized markov chains for next-basket recommendation,'' in \emph{WWW}, 2010, pp. 811--820.

\bibitem{r5}
S.~Wan, Y.~Lan, P.~Wang, J.~Guo, J.~Xu, and X.~Cheng, ``Next basket recommendation with neural networks,'' in \emph{RecSys (Poster)}, 2015.

\bibitem{r15}
P.~Wang, J.~Guo, Y.~Lan, J.~Xu, S.~Wan, and X.~Cheng, ``Learning hierarchical representation model for next-basket recommendation,'' in \emph{SIGIR}, 2015, pp. 403--412.

\bibitem{r1}
F.~Yu, Q.~Liu, S.~Wu, L.~Wang, and T.~Tan, ``A dynamic recurrent model for next basket recommendation,'' in \emph{SIGIR}, 2016, pp. 729--732.

\bibitem{r12}
H.~Hu and X.~He, ``Sets2{S}ets: Learning from sequential sets with neural networks,'' in \emph{SIGKDD}, 2019, pp. 1491--1499.

\bibitem{r13}
D.-T. Le, H.~W. Lauw, and Y.~Fang, ``Correlation-sensitive next-basket recommendation,'' in \emph{IJCAI}, 2019, pp. 2808--2814.

\bibitem{r16}
T.~Bai, J.-Y. Nie, W.~X. Zhao, Y.~Zhu, P.~Du, and J.-R. Wen, ``An attribute-aware neural attentive model for next basket recommendation,'' in \emph{SIGIR}, 2018, pp. 1201--1204.

\bibitem{wang2020intention}
S.~Wang, L.~Hu, Y.~Wang, Q.~Z. Sheng, M.~Orgun, and L.~Cao, ``Intention {N}ets: Psychology-inspired user choice behavior modeling for next-basket prediction,'' in \emph{AAAI}, 2020, pp. 6259--6266.

\bibitem{wang2021intention2basket}
S.~Wang, L.~Hu, Y.~Wang, Q.~Z. Sheng, M.~Orgun, and et~al., ``Intention2{B}asket: A neural intention-driven approach for dynamic next-basket planning,'' in \emph{IJCAI}, 2020, pp. 2333--2339.

\bibitem{hu2020modeling}
H.~Hu, X.~He, J.~Gao, and Z.~Zhang, ``Modeling personalized item frequency information for next-basket recommendation,'' in \emph{SIGIR}, 2020, pp. 1071--1080.

\bibitem{r14}
G.~Faggioli, M.~Polato, and F.~Aiolli, ``Recency aware collaborative filtering for next basket recommendation,'' in \emph{UMAP}, 2020, pp. 80--87.

\bibitem{wang2020hierarchical}
S.~Wang, L.~Cao, L.~Hu, S.~Berkovsky, X.~Huang, L.~Xiao, and W.~Lu, ``Hierarchical attentive transaction embedding with intra-and inter-transaction dependencies for next-item recommendation,'' \emph{IEEE Intelligent Systems}, vol.~36, no.~4, pp. 56--64, 2020.

\bibitem{r26}
M.~Ludewig and D.~Jannach, ``Evaluation of session-based recommendation algorithms,'' \emph{User Modeling and User-Adapted Interaction}, vol.~28, no.~4, pp. 331--390, 2018.

\bibitem{r7}
S.~Wang, L.~Hu, Y.~Wang, L.~Cao, Q.~Z. Sheng, and M.~Orgun, ``Sequential recommender systems: Challenges, progress and prospects,'' in \emph{IJCAI}, 2019, pp. 6332--6338.

\bibitem{r9}
H.~Fang, G.~Guo, D.~Zhang, and Y.~Shu, ``Deep learning-based sequential recommender systems: Concepts, algorithms, and evaluations,'' in \emph{Proceedings of International Conference on Web Engineering}, 2019, pp. 574--577.

\bibitem{guo2021sequential}
W.~Guo, S.~Wang, W.~Lu, H.~Wu, Q.~Zhang, and Z.~Shao, ``Sequential dependency enhanced graph neural networks for session-based recommendations,'' in \emph{DSAA}, 2021, pp. 1--10.

\bibitem{hu2017diversifying}
L.~Hu, L.~Cao, S.~Wang, G.~Xu, J.~Cao, and Z.~Gu, ``Diversifying personalized recommendation with user-session context,'' in \emph{IJCAI}, 2017, pp. 1858--1864.

\bibitem{wang2022exploiting}
N.~Wang, S.~Wang, Y.~Wang, Q.~Z. Sheng, and M.~A. Orgun, ``Exploiting intra- and inter-session dependencies for session-based recommendations,'' \emph{World Wide Web Journal}, vol.~25, no.~1, pp. 425--443, 2022.

\bibitem{r24}
M.~Beladev, L.~Rokach, and B.~Shapira, ``Recommender systems for product bundling,'' \emph{Knowledge-Based Systems}, vol. 111, no.~C, pp. 193--206, 2016.

\bibitem{r25}
J.~Bai, C.~Zhou, J.~Song, X.~Qu, W.~An, Z.~Li, and J.~Gao, ``Personalized bundle list recommendation,'' in \emph{WWW}, 2019, pp. 60--71.

\bibitem{hao2020p}
J.~Hao, T.~Zhao, J.~Li, X.~L. Dong, C.~Faloutsos, Y.~Sun, and W.~Wang, ``P-companion: {A} principled framework for diversified complementary product recommendation,'' in \emph{CIKM}, 2020, pp. 2517--2524.

\bibitem{chen2019matching}
L.~Chen, Y.~Liu, X.~He, L.~Gao, and Z.~Zheng, ``Matching user with item set: Collaborative bundle recommendation with deep attention network,'' in \emph{IJCAI}, 2019, pp. 2095--2101.

\bibitem{r19}
H.~Fang, D.~Zhang, Y.~Shu, and G.~Guo, ``Deep learning for sequential recommendation: Algorithms, influential factors, and evaluations,'' \emph{ACM Transactions on Information Systems}, vol.~39, no.~1, pp. 1--42, 2020.

\bibitem{r10}
R.~Guidotti, G.~Rossetti, L.~Pappalardo, F.~Giannotti, and D.~Pedreschi, ``Market basket prediction using user-centric temporal annotated recurring sequences,'' in \emph{ICDM}, 2017, pp. 895--900.

\bibitem{ariannezhad2022recanet}
M.~Ariannezhad, S.~Jullien, M.~Li, M.~Fang, S.~Schelter, and M.~de~Rijke, ``Re{CAN}et: A repeat consumption-aware neural network for next basket recommendation in grocery shopping,'' in \emph{SIGIR}, 2022.

\end{thebibliography}

\vspace{12pt}
\color{red}

\end{CJK}
\end{document}